\documentclass[aps,prd,superscriptaddress,nofootinbib,eqsecnum,twocolumn]{revtex4-1}

\pdfoutput=1

\usepackage{graphicx}  % needed for figures
\usepackage{dcolumn}   % needed for some tables
\usepackage{amssymb}   % for math
\usepackage{xcolor}
\usepackage{graphics} 
\usepackage{hyperref}
\usepackage{amsfonts}
\usepackage{latexsym}
\usepackage{slashed}
\usepackage{amsmath}
\usepackage{amsfonts}
\usepackage{amssymb}
\usepackage{dsfont}
\usepackage{bm}  % To make greek letters in bold face {\bm \Phi}
\usepackage[normalem]{ulem} %tabela
\usepackage{amsmath,mathtools,mathrsfs,bbm,amssymb} 
\usepackage{bbold}
\usepackage{subfigure}
\usepackage{array}
\usepackage{wrapfig}
\usepackage{multirow}
\usepackage{tabularx}
\usepackage{float}
\usepackage{array}
\usepackage{booktabs}
\usepackage{commath}
\usepackage{graphicx}
\usepackage{epstopdf}
\usepackage{amsmath}
\usepackage{amsfonts}
\usepackage{amssymb}
\usepackage{dsfont}
\usepackage{latexsym}
\usepackage{hyperref}
\usepackage[english]{babel}
\usepackage[utf8]{inputenc}
\usepackage[colorinlistoftodos]{todonotes}
\usepackage{xcolor}
\usepackage{slashed}
\usepackage{bm}  % To make greek letters in bold face {\bm \Psi}
\usepackage[normalem]{ulem} %tabela
\usepackage{amsmath,mathtools}
\usepackage{caption}
\usepackage{subcaption}
\usepackage{float}
\usepackage{MnSymbol}
\useunder{\uline}{\ul}{}
\usepackage{hyperref}
\begin{document}

\title{Dissipative stabilization of Ostrogradsky modes in non-equilibrium field theory}
\author{Y. M. P. Gomes}
\email{ymuller@cbpf.br}
 \affiliation{Centro Brasileiro de Pesquisas F\'isicas, Rua Dr. Xavier Sigaud 150, Urca, CEP: 22290-180, Rio de Janeiro-RJ, Brazil}

\begin{abstract}
In this work, we investigate higher-derivative quantum field theories and the problem of Ostrogradsky instability within an open-system Keldysh-Lindblad framework. Coupling the ghost sector to dissipative baths generates non-perturbative effective masses and dissipative widths through self-consistent gap equations. Above a critical coupling, the nonequilibrium dynamics develops bifurcated dissipative branches, signaling the emergence of a dissipative phase transition and a nontrivial critical structure in parameter space. We find that the resulting dissipative dynamics can suppress ghost excitations through two distinct mechanisms: in one branch, a large dynamically generated effective mass preserves a quasiparticle-like excitation, while in the second branch, strong dissipative broadening destroys the quasiparticle character through overdamped dynamics. Our results suggest that dissipative effects may provide a nonequilibrium mechanism for the spectral suppression of Ostrogradsky ghosts. The comparison with the healthy sector indicates that the stabilization mechanism is intrinsically tied to the ghost-like spectral structure.

\end{abstract}

\maketitle

\section{Introduction}
 
There have been many proposals to employ ghosts, excitations whose propagators carry a negative sign, to address some of the most difficult phenomena in the universe. For example, Steele demonstrates how these ghosts are connected to effective theories of gravity \cite{Stelle77}.  
Terms in the Lagrangian as $R^2$, $R_{\mu \nu}^2$, or $\phi \square^2 \phi$ introduces fourth order derivatives and new massive modes in the model that can propagate as ghosts. However, these modes' behavior can bring with it the quantum version of the Ostrogradsky instability, breaking the unitarity of the theory and destroying the probabilistic interpretation of the quantum theory itself \cite{Julve78,Tomboulis80,Bender08,Smilga05,Smilga17,Kaparulin14}. More recently, the problem has resurged with new approaches from the analysis of physical bound states of ghosts \cite{Asorey26, Modesto16, Liu23}, and by the analogy between ghosts in gravity and quantum Chromodynamics (QCD) \cite{Debrito24}. 

%Lacuna na Literatura
Interestingly, the comprehension of those systems with higher derivatives in the non-equilibrium regime is still underrated. Particularly, the Keldysh-Schwinger is a natural and powerful tool to examine out-of-equilibrium QFT \cite{Kadanoff62, Schwinger60,Keldysh64}. 
The nonequilibrium quantum phenomena have become increasingly relevant for the understanding of several modern physical systems, revealing structures and dynamical effects that cannot be captured within equilibrium approaches alone. This perspective has motivated intense investigations in areas such as relativistic quantum transport and spin kinetics \cite{Gao19}, quantum time crystals \cite{Wilczek12,Lazarides17}, and anomalous chiral transport phenomena in high-energy nuclear processes \cite{Kharzeev16}. These developments indicate that dissipative and nonequilibrium effects may play a fundamental role in the emergence of novel quantum phases and collective behaviors \cite{Martin59,Berges04,Kamenev10,Kamenev23,Thompson23,Sieberer16}. Particularly, for some configurations, the effective dynamics of the subsystem can be described by a Markovian master equation. In such cases, the quantum systems run out of the unitary dynamics dictated by the Hamiltonian formalism, but via the Lindblad master equation, which preserves the probabilistic interpretation, despite the lack of unitarity \cite{Gorini76,Lindblad76}.

By use of the Keldysh-Schwinger formalism, we reach the generalization of the Boltzmann equation for higher derivative models, more accurately for scalar models with four derivatives. By decomposition into healthy and ghost components, we are able to understand the properties of the model mode deeply. Analyzing the stability of the system when the ghost sector is present, the resulting Keldysh action can be derived, and characteristic Dyson-Schwinger equations within Boltzmann equations, which rule the hydrodynamic behavior of the system is found. 

Going further, we investigate the system when coupled with a massive environment described effectively by the Lindblad equation.  By use of Keldysh formalism, we derive the effective action for the open system and analyze its non-perturbative and hydrodynamical properties. By the introduction of quadratic jump operators for the healthy and ghost sectors, and via the 2-PI formalism, we find self-consistent gap equations that rule the effective dynamics of the excitations. 
The existence of condensate solutions can renormalize the masses and possibly stabilize the ghost sector. In this work, we investigate whether the quantum dissipation is able to provide an effective dissipative suppression of Ostrogradsky modes, and our findings can shed light on similar problems, such as quantum theories for gravity and out-of-equilibrium cosmological scenarios. 

The organization of the present work is the following. In Sec. II, we discuss the non-equilibrium dynamics of the closed higher-derivative theory within the Keldysh framework, emphasizing the absence of a collision term and the difficulties in defining a genuine kinetic notion of stability in the free-streaming regime. In Sec. III, we introduce the open-system description and discuss how dissipative effects emerge from the coupling to external environments, leading to effective spectral stabilization. In Sec. IV, we present the quartic higher-derivative model, its Ostrogradsky structure, and the corresponding Keldysh-Lindblad effective action. We derive the 2PI gap equations governing the effective mass and dissipative width of the ghost sector and analyze their non-trivial solutions numerically, revealing the emergence of bifurcated dissipative branches and a critical structure in parameter space. Finally, the appendices contain the derivation of the effective Lindblad action from microscopic baths and the exact evaluation of the gap integrals used throughout the work.

%%%%%%%%%%%%%%%%%%%%%%%%%%%%%%%%%%%%%%%%%%%%%%%%%%%%%%%%%%%%%%%%%%%%%%%

\section{Open-system dynamics and effective description}

We now investigate how dissipative effects modify the dynamics of Ostrogradsky modes in a nonequilibrium setting. 
Rather than treating the higher-derivative theory as an isolated system, we consider an effective open-system description in which the ghost sector interacts with external environments. 
Within the Keldysh--Lindblad framework, such coupling generates imaginary self-energy contributions in the retarded propagator, leading to dissipative broadening and nontrivial spectral dynamics.

The central question is whether dissipation can dynamically suppress the pathological growth associated with ghost excitations. 
To address this issue, we analyze self-consistent gap equations for the effective mass and dissipative width generated by quadratic Lindblad interactions. 
As we will show, the resulting nonequilibrium dynamics admits nontrivial dissipative branches above a critical coupling, leading to a bifurcated phase structure in parameter space. For the sake of completeness, an equivalent coupling of the healthy sector with an external environment is also analyzed.

%------------------------------------------------------------------------------------------------
\subsection{Higher derivative model: Quartic case}
In this section, one looks to a field equivalent to the Ostrogradsky model, which can be described via the following Lagrangian:
\begin{equation}
    \mathcal{L} = \phi(x) \left( \beta^4 + \alpha^2 \square + \square^2\right) \phi(x) ~~. 
\end{equation}
and, thus, in momentum space one has $\mathcal{O}(p^2)=\left(p^4 - \alpha^2 p^2 + \beta^4\right)$.
Therefore, can be shown that the poles of the propagator are given by:
\begin{equation}
    p^2 = \frac{\alpha^2}{2} +\frac{\ell }{2}\sqrt{\alpha^4 - 4 \beta^4} = M_{\ell}^2~~,~~ \ell=\pm 1~~.
\end{equation}
Importantly, if $\alpha^2<2 \beta^2$, the masses become complex (conjugate one relative to the other). The dispersion relation is given by:
\begin{equation}
    \omega_{ \ell}({\bf p}) = \pm \sqrt{{\bf p}^2 + M_{\ell}^2}~~,~~ \ell = \pm 1~~.
\end{equation}

The quartic operator $\mathcal{O}(\square) = \beta^4 +\alpha^2\square + \square^2$ corresponds to a Lagrangian with fourth-order time derivatives. Applying the Ostrogradsky method, we introduce two canonical pairs $(q_1 = \phi, \pi_1)$ and $(q_2 = \dot{\phi}, \pi_2)$, yielding the Hamiltonian
\begin{eqnarray}\nonumber
\mathcal{H} &=& \left[ \frac{\pi_2^2}{2} + \pi_1 q_2 + \frac{\alpha^2}{2} q_2^2 - \frac{\alpha^2}{2} (\nabla q_1)^2 - \frac{\beta^4}{2} q_1^2 +\right.\\
&&\left. + \pi_2 \nabla^2 q_1 - \frac{1}{2} (\nabla^2 q_1)^2 \right].
\end{eqnarray}
The key feature is the linear term $\pi_1 q_2$, which makes the Hamiltonian unbounded from below. The Hamilton equation reads:
\begin{eqnarray}
&&\dot{q}_1 = q_2 ~~,\\
&&\dot{q}_2 = \pi_2 + \nabla^2 q_1 ~~,\\
&&\dot{p}_1 = \nabla^2 \pi_2 + \beta^4 q_1 +\alpha^2 \nabla^2 q_1 ~~,\\
&&\dot{p}_2 = -\pi_1 +\alpha^2 q_2~~.
\end{eqnarray}
The diagonalized degrees of freedom are the following :
\begin{equation}
    \Phi({\bf p}, t) = q_2({\bf p}, t) + i \omega_{-}({\bf p }) q_1({\bf p}, t)~~,
\end{equation}
the healthy sector and
\begin{equation}
    \tilde{\Phi}({\bf p}, t) = q_2({\bf p}, t) + i \omega_{+}({\bf p }) q_1({\bf p}, t)~~,
\end{equation}
the ghost sector. The original fields are expressed as:
\begin{equation}
    q_1 = \frac{\tilde{\Phi} - \Phi}{i(\omega_+ - \omega_-)}~~,~~ q_2 = \frac{\omega_+ \Phi - \omega_- \tilde{\Phi}}{\omega_+ - \omega_-}~~.
\end{equation}
The physical field $\phi=q_1$ is thus a superposition of the healthy and ghost modes; the ghost cannot be eliminated by a local field redefinition. Upon such diagonalization, the Hamiltonian describes two modes with energies $\omega_\ell$, and an action can be rewritten as follows:
\begin{equation}
    \mathcal{S} = \frac{1}{2 \Delta} \int d^4x \Big[ \Phi (\square + M_-^2) \Phi - \tilde{\Phi} (\square + M_+^2) \tilde{\Phi} \Big]~~,
\end{equation}
with $\Delta = \sqrt{\alpha^4 - 4\beta^4}$. Importantly, the massive mode contributes with a negative sign, i.e., $H = \sum_{\mathbf{p}} (\omega_-a_{\mathbf{p}}^\dagger a_{\mathbf{p}} - \omega_+ b_{\mathbf{p}}^\dagger b_{\mathbf{p}})$. This negative weight is the microscopic origin of the ghost, which manifests in the spectral function as $\rho(p) \propto \delta(p^2 - M_-^2) - \delta(p^2 - M_+^2)$.

%-------------------------------------------------------

\subsection{Keldysh construction for $\Phi$}
In terms of the healthy and ghosts d.o.f.s,  the Keldysh effective action is:
\begin{equation}
    \mathcal{S}_K = \frac{1}{2 \Delta} \int d^4x \Big[ \Phi_q (\square + M_-^2) \Phi_c - \tilde{\Phi}_q (\square + M_+^2) \tilde{\Phi}_c \Big] ~~.
\end{equation}
Both sectors remain free, while the ghost ($\tilde{\Phi}$) experiences a minus sign as expected. The retarded propagators satisfy:
\begin{align}
    (\square_x + m_-^2) G^R_\Phi(x, y) &= +2 \Delta \, \delta^{(4)}(x - y), \\
    (\square_x + m_+^2) G^R_{\tilde{\Phi}}(x, y) &= -2 \Delta \, \delta^{(4)}(x - y),
\end{align}
Using the Wigner representation, the Keldysh equations reduce to the Vlasov equations:
\begin{align}
    p^\mu \partial_\mu F_\Phi(X, p) &= 0 \quad \text{(healthy sector)}, \\
    p^\mu \partial_\mu F_{\tilde{\Phi}}(X, p) &= 0 \quad \text{(ghost sector)}.
\end{align}
The negative kinetic residue of the ghost cancels out in the transport equation, leaving both sectors with the same free-streaming behavior. The Ostrogradsky instability manifests only in the sign of the ghost propagator. When interactions are present, new features can appear, either through the collision term or through the fluctuation-dissipation relation. 

%--------------------------------------------------------%--------------------------------------------------------%--------------------------------------------------------

\subsection{Out-of-equilibrium ghosts }

The Lindblad structure derived in the sequel should be understood as an effective open-system description obtained after integrating environmental degrees of freedom. Therefore, the dissipative dynamics is not assumed to define a fundamental microscopic theory for the ghost sector, but rather an emergent non-equilibrium effective description valid within a restricted dynamical regime. If we extend the dynamics os the system to an open system described by the Lindblad equation given by:
\begin{equation}\label{lind1}
    \partial_t{\rho} = -i[H, \rho] + \sum_n \gamma_n \left( L_n \rho L_n^\dagger - \frac{1}{2} \{ L_n^\dagger L_n, \rho \} \right)~~~,
\end{equation}
where $L_n$ is the jump operator, which couples the system with the environment, and $\gamma_n$ are the respective coupling constants. The jump operators and the resulting coupling constants emerge after integrating the environment degrees of freedom properly (See Appendix \ref{AppI}). Thus, Eq. \eqref{lind1} can be seen as an effective equation for the system. 
%\subsection{Keldysh Action with Ghost Jump Operator}

For a Lindblad jump operator $L =  \tilde{\Phi}^2$ acting on the ghost sector, the Keldysh effective action is:
\begin{eqnarray}\label{ghostself}\nonumber
    &&\mathcal{S}_K = \frac{1}{2 \Delta} \int d^4x \Big[ \Phi_q (\square + M_-^2) \Phi_c - \tilde{\Phi}_q (\square + M_+^2) \tilde{\Phi}_c \Big] +\\
    &&\hspace{.5cm}+ 4 i \gamma \int d^4x \, \tilde{\Phi}_c^2 \tilde{\Phi}_q^2 ~~.%+ 4 i \zeta \int d^4x \, \left({\Phi}_c \tilde{\Phi}_q+\tilde{\Phi}_c {\Phi}_q\right)^2 ~.\\
\end{eqnarray}
The healthy sector $\Phi$ remains free. 
For $\gamma \neq 0$, the 2PI effective action separates into a free healthy sector and an interacting ghost sector:
\begin{equation}
\begin{aligned}
    \Gamma[\mathbf{G}_\Phi, \mathbf{G}_{\tilde{\Phi}}] &= \frac{i}{2} \operatorname{Tr} \ln \mathbf{G}_\Phi^{-1} + \frac{i}{2} \operatorname{Tr} (\mathbf{G}_{0,\Phi}^{-1} \mathbf{G}_\Phi) \\
    &\quad + \frac{i}{2} \operatorname{Tr} \ln \mathbf{G}_{\tilde{\Phi}}^{-1} + \frac{i}{2} \operatorname{Tr} (\mathbf{G}_{0,\tilde{\Phi}}^{-1} \mathbf{G}_{\tilde{\Phi}}) \\
    &\quad +\Gamma_{Hartree}[\mathbf{G}_{\tilde{\Phi}}] + \Gamma_{\text{Sunset}}[\mathbf{G}_{\tilde{\Phi}}].
\end{aligned}
\end{equation}
with
\begin{eqnarray}\nonumber 
\Gamma_2^{H}[G]
&=&
\gamma \int d^4x \,
\Big[
\big(\mathrm{Tr}[\sigma_x \mathbf{G}_{\tilde{\Phi}}(x,x)]\big)^2
+\\
&&
2\,\mathrm{Tr}\!\left(\sigma_x \mathbf{G}_{\tilde{\Phi}}(x,x)\sigma_x \mathbf{G}_{\tilde{\Phi}}(x,x)\right)
\Big]~~.
\end{eqnarray}
The sunset term $\Gamma_{\text{Sunset}}$ contains the two-loop diagrams generated by the quartic vertex, and it is given by:

\begin{eqnarray}\nonumber
 &&\Gamma_{\text{Sunset}}[\mathbf{G}_{\tilde{\Phi}}]
=\\
&&=
4\gamma^2
\int d^4x\, d^4y \,
\Bigg[
\left(
\mathrm{Tr}\!\left[
\sigma_x \mathbf{G}_{\tilde{\Phi}}(x,y)\sigma_x \mathbf{G}_{\tilde{\Phi}}(y,x)
\right]
\right)^2+
\nonumber\\\nonumber
&& +2\,\mathrm{Tr}\!\left[
\sigma_x \mathbf{G}_{\tilde{\Phi}}(x,y)\sigma_x \mathbf{G}_{\tilde{\Phi}}(y,x)
\sigma_x \mathbf{G}_{\tilde{\Phi}}(x,y)\sigma_x \mathbf{G}_{\tilde{\Phi}}(y,x)
\right]
\Bigg]~~.\\
\end{eqnarray}

Interestingly, for the self-interaction shown in Eq. \eqref{ghostself}, the sunset term vanishes exactly. The cancellation follows from the causal support of the retarded and advanced propagators in Keldysh space, which forbids non-vanishing contributions from the corresponding two-loop structures. From the 2PI effective action, the retarded and Keldysh self-energies of the ghost field $\tilde{\Phi}$ are:
\begin{eqnarray}\nonumber
\Sigma^{R}_{\tilde{\Phi}}(x,y)
&=&
4\,i\gamma\,\delta^{(4)}(x-y)\,
\Big[
{G}_{\tilde{\Phi}}^R(x,x)+3{G}_{\tilde{\Phi}}^A(x,x)
\Big]~~,~~\\
\\[1ex]
\Sigma^{K}_{\tilde{\Phi}}(x,y)
&=&
8\,i\gamma\,\delta^{(4)}(x-y)\,
{G}_{\tilde{\Phi}}^K(x,x)~~.
\end{eqnarray}
The Hartree term provides a local mass shift for the ghost sector, and the healthy mode remains free.
The Boltzmann equation now reads:
\begin{align}\label{boltz1}
    p^\mu \partial_\mu F_\Phi(X, p) &= 0 ~~, \\
    p^\mu \partial_\mu F_{\tilde{\Phi}}(X, p) &= C(X,p)~~
\end{align}

with
\begin{eqnarray}\nonumber
C(X,p)
&=& -8\gamma\left( \,F_{\tilde{\Phi}}(X,p)\,\lambda_{\tilde{\Phi}}(X,X)-  G^K(X,X) \right)~~,\\
\end{eqnarray}
with $\lambda_{\tilde{\Phi}}(x,y)=i\big[{G}_{\tilde{\Phi}}^R(x,y)-{G}_{\tilde{\Phi}}^A(x,y)\big]$ the spectral function of the ghost and $\lambda_{\tilde{\Phi}}(X,X)= \lambda_{\tilde{\Phi}}(x,y)|_{x=y=X}$ (same for $G^K(X,X)$). 
Defining $\tau(X) = {G}_{\tilde{\Phi}}^R(X,X)+{G}_{\tilde{\Phi}}^A(X,X)$ and assuming constant configurations $\tau,\rho$, the retarded green function for the ghost field reads:
\begin{equation}
    G^R_{\tilde{\Phi}} (p) = - \frac{\Delta}{p^2 - M_+^2 - 4 \gamma \rho + 8 i \gamma \tau}~~,
\end{equation}

Through introduction of Euclidean cutoff $\Lambda$, we reach \ref{AppII}:

\begin{align}
    \tau &= - \frac{\Delta}{8\pi^2} \left[ \Gamma^2 \ln L + M_R^2 \, \theta \right], \\
    \rho &=  \frac{\Delta}{8\pi^2} \left[ \Lambda^2 - M_R^2 \ln L + \Gamma^2 \, \theta \right].
\end{align}
with:
\begin{align}
    L &= \sqrt{ \frac{ (\Lambda^2 + M_R^2)^2 + \Gamma^4 }{ (M_R^2)^2 + \Gamma^4 } }, \\
    \theta &= \arctan\left( \frac{\Gamma^2}{\Lambda^2 + M_R^2} \right) - \arctan\left( \frac{\Gamma^2}{M_R^2} \right)~~,
\end{align}
with $M_R^2 = M_+^2 + 4 \gamma \rho$ and $\Gamma=8 \gamma \tau$. Lastly, we need to look at the Boltzmann equation. From eq. \ref{boltz1}, we can rewrite the collision term as follows:
\begin{eqnarray}\nonumber
&&C[F](X,p) =\\\nonumber
&&=-8\gamma\int \frac{d^4q}{(2\pi)^4}\,\lambda_{\tilde{\Phi}}(X,q)\left[F_{\tilde{\Phi}}(X,q)-F_{\tilde{\Phi}}(X,p)\right]~~,\\    
\end{eqnarray}
with $\lambda_{\tilde{\Phi}}(X,q)$ the spectral function in Wigner space. The following quantity is conserved:
\begin{equation}
   J^\mu = \int \frac{d^4q}{(2\pi)^4}\,\lambda_{\tilde{\Phi}}(X,q) p^\mu F(X,p)~,~\partial_\mu J^\mu = 0~~.
\end{equation}

It is worth emphasizing that the same construction can be applied to the healthy sector by considering a quadratic jump operator $L' =  {\Phi}^2$, and the gap equations remain, with the simple replacement $M_+\to M_-$ combined with $\rho \to - \rho$.
Although the healthy sector can be treated by the same Keldysh--Lindblad construction, its self-consistent mass shift has the opposite sign, $M_{R,-}^2=M_-^2-4\gamma\rho,
$ in contrast with the ghost sector, $M_{R,+}^2=M_+^2+4\gamma\rho$.
Therefore, the dissipative bifurcation is not obtained by a simple replacement $M_+\to M_-$. The two sectors display qualitatively different phase structures, with the ghost sector exhibiting a positive Hartree feedback that can drive mass generation, while the healthy sector experiences an opposite mass renormalization. In the following section, we discuss the results for both cases.

%------------------------------------------------------------------
\section{Results}
\subsection{Ghost sector}
Analyzing the ghost sector, if we look to the sector $\tau=0$, the non-trivial gap equation reads:
\begin{equation}
M_R^2=M_+^2+ \frac{\gamma \Delta}{2\pi^2} \left[ \Lambda^2 - M_R^2 \ln \frac{ (\Lambda^2 + M_R^2) }{ M_R^2 } \right]
\end{equation}
The above equation is equal to the Nambu-Jona-Lasinio (NJL) gap equation with $G = \Delta \gamma$. Thus, in the phase where $\tau=0$, the ghost mode gains an effective mass $M_R$ due to a dissipative phase transition (DPT), and remains stable. Going further, we can look to the sector $\rho=0$. In this case, we find

\begin{equation}
    \Gamma = - \frac{\gamma\Delta}{\pi^2} \left[  \left( \frac{\Lambda^2 \Gamma^2}{M_+^2}\right)+ \left( \frac{ \Gamma^4 +M_+^4\, }{M_+^2}\right)\theta  \right],
\end{equation}
which is automatically fulfilled for $\Gamma=0$. Nonetheless, a non-trivial solution can exist, and we can find, for instance 
\begin{eqnarray}\label{tauappr}
\Gamma
&\simeq&
-\frac{\pi^2}{\gamma\Delta\,\mathcal{C}} ~~,~~ \frac{1}{\gamma \Delta \mathcal{C}} \ll M_+~~,
\end{eqnarray} 
with $\mathcal{C}= \ln\left(1+\frac{\Lambda^2}{M_+^2}\right)
-\frac{\Lambda^2}{\Lambda^2+M_+^2} >0$.  Thus, for a sufficiently strong $\gamma$, $\tau$ can assume a non-null and negative value given approximately by Eq. \eqref{tauappr}.  Therefore, we find non-trivial solutions $\Gamma\neq 0$ for sufficiently strong couplings. For the retarded propagator convention adopted here, the condition $\Gamma<0$ shifts the poles toward damped configurations in the lower-half complex plane, suppressing the exponential growth associated with the ghost excitation.

\begin{figure}
    \centering
    \includegraphics[width=1\linewidth]{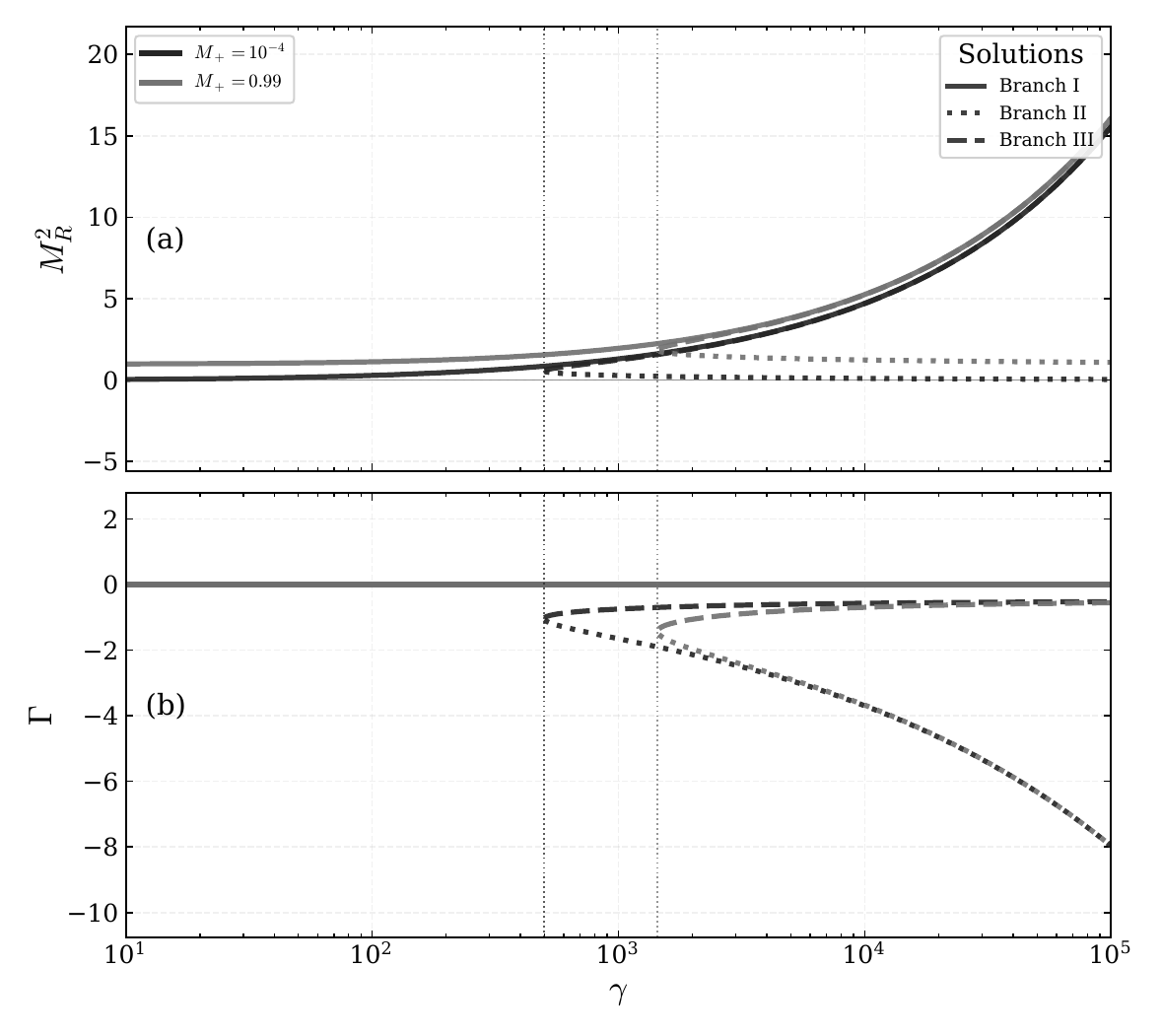}
    \caption{(a) Effective mass squared $M_R^2 $ of the ghost sector, in units of $\Lambda$, as a function of $z = \log_{10}(\gamma )$, for fixed $\Delta = 0.1$ and several values of the bare mass $M_+ $. The NJL-like branch ($\Gamma = 0$) exists for all $z$, while beyond a critical coupling $z_c$ two additional bifurcated branches appear (upper and lower). The upper branch grows with increasing $z$, while the lower branch remains finite and eventually saturates. The bifurcation point $z_c$ shifts to larger values as $M_+$ increases. (b) Dissipative width $\Gamma / \Lambda^2$ of the ghost sector as a function of $z = \log_{10}(\gamma / \Lambda^2)$, for $\Delta = 0.1$ and the same values of $M_+ / \Lambda$ as in Fig.~1. The trivial solution $\Gamma = 0$ is present for all $z$. At the bifurcation point $z_c$, two non-trivial branches with $\Gamma \neq 0$ emerge simultaneously. In both branches $\Gamma$ remains negative, corresponding to a damped (stable) spectral configuration. The magnitude of $\Gamma$ grows with $z$ in the upper branch, while it remains small in the lower branch, indicating distinct dissipative regimes.}
    \label{fig1}
\end{figure}

\begin{figure}
    \centering
    \includegraphics[width=1\linewidth]{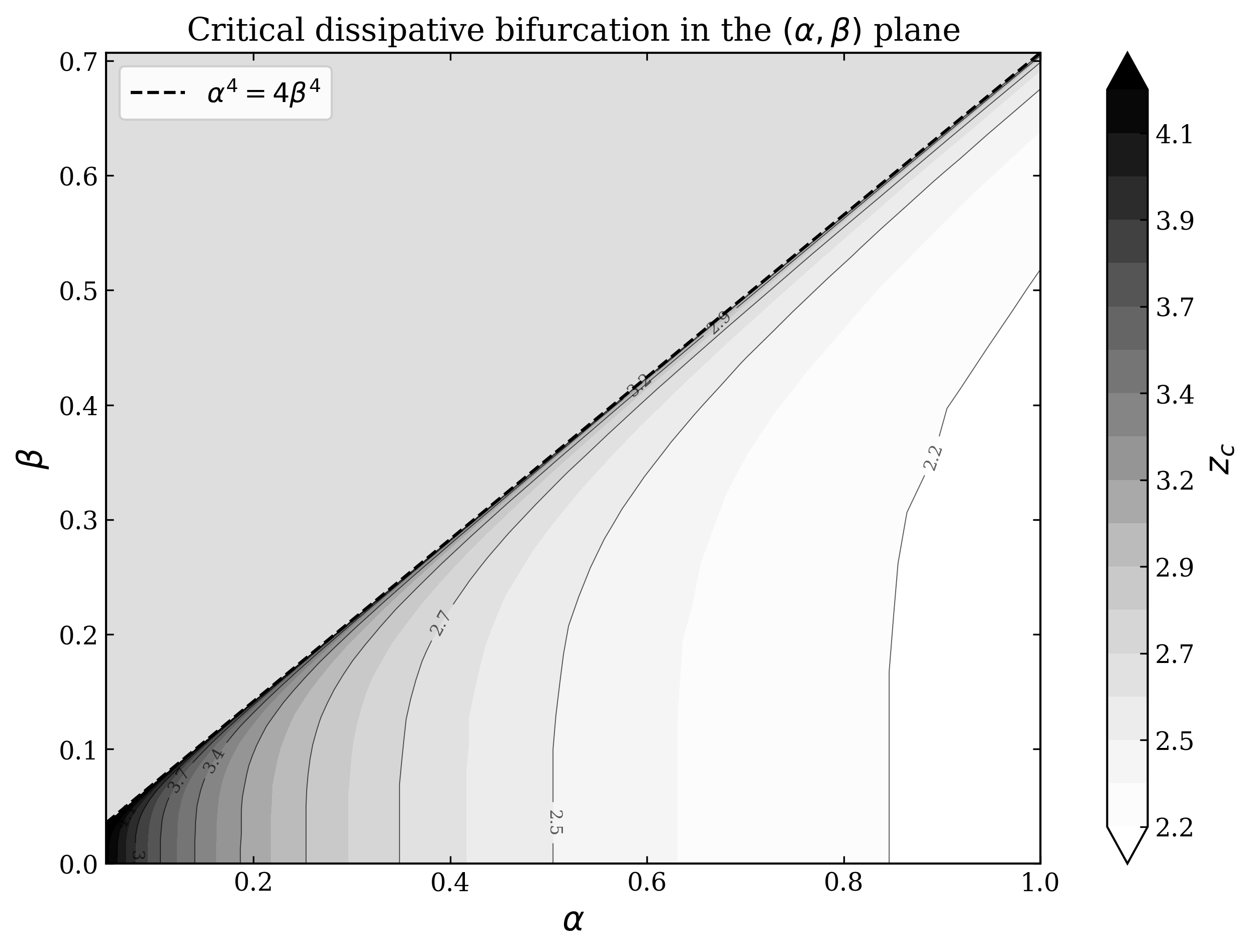}
   \caption{
Phase diagram in the $(\alpha,\beta)$ plane showing contour lines of the critical coupling
$z_c=\log_{10}\gamma_c$ is associated with the onset of the bifurcated dissipative regime. 
The shaded region corresponds to the non-physical domain 
$\alpha^4 < 4\beta^4$, where the spectral splitting 
$\Delta=\sqrt{\alpha^4-4\beta^4}$ becomes imaginary. 
The dashed curve 
$\alpha^4=4\beta^4$ marks the degenerate limit $\Delta=0$, where the healthy and ghost modes acquire the same mass. }
    \label{fig2}
\end{figure}

By numerical analysis, we can see that the gap equations exhibit a non-trivial solution structure, with multiple branches for $\Gamma \neq 0$ emerging beyond a critical coupling depending on both $M_+$ and $\Delta$. A bifurcation occurs at a well-defined value $z_c = \log_{10}\gamma_c$, where two non-trivial branches appear simultaneously. Importantly, one branch of $M_R$ and $\Gamma$ grows with z, while the other remains small and finite as $z \to \infty$. In the branch where $|\Gamma|$ dominates over $M_R$, the spectral function becomes strongly broadened, indicating the loss of a well-defined quasiparticle description for the ghost excitation.
Moreover, the location of the bifurcation depends on the mass parameter $M_+$, shifting to larger $z$ as $M_+$ increases. On the other hand, the critical value $z_c$ decreases as $\Delta$ increases. Since $\Delta$ represents the mass difference between the healthy and ghost sectors, this separation favours the DPT and the emergence of a new order parameter. This behavior indicates that the spectral separation between the healthy and ghost sectors plays a crucial role in triggering the dissipative instability. 

The NJL-like solution $M_R \neq 0$ with $\Gamma=0$ remains present for all $z$, but becomes non-unique beyond $z_c$. Beyond the critical point, the dissipative solutions coexist with the NJL-like branch, revealing the non-uniqueness of the self-consistent nonequilibrium vacuum. The quantity $\Gamma $ provides a natural order parameter, vanishing in the trivial phase and becoming finite in the bifurcated regime, remaining negative in both branches of the bifurcation. 
The transition is therefore characterized by the spontaneous emergence of a finite dissipative width, separating a trivial quasiparticle-like regime from a genuinely dissipative nonequilibrium phase.
The non-trivial branches are clearly separated and display distinct magnitudes of $|\Gamma|$, indicating a structured hierarchy of solutions.
These results point to the dynamical generation of a non-trivial scale, which can be interpreted as the onset of a new phase. The coexistence of multiple self-consistent branches suggests a nontrivial landscape of dissipative fixed points in parameter space.
The question of the stability of the different branches remains open and requires a separate analysis.

At $\alpha = \sqrt{2}\beta$ the modes are degenerated, and the accumulation of contour lines near this boundary indicates that the critical coupling increases sharply as the degeneracy limit is approached. For $\alpha < \sqrt{2}\beta$ both modes become unstable (shaded region of Fig. \ref{fig2}). 
%Since the present framework should be understood as an effective nonequilibrium field theory, the ultraviolet cutoff $\Lambda$ defines the natural scale delimiting the validity regime of the dissipative description.

%%%%%%%%%%%%%%%%%%%%%%%%%%%%%%%%%%%%%%%%%%%%
\subsection{Healthy sector}
Going further, in the healthy sector, the effective mass $M_R^2$ and width $\Gamma$ will behave very differently. Since the Hartree self-energy carries the same residue structure as the corresponding propagator, the resulting gap equations have the same functional form. Therefore, the dynamical generation of an effective mass and a spectral width is a generic feature of the quadratic Lindblad self-interaction, while its interpretation as a stabilization mechanism is specific to the ghost sector, where the dissipative width regulates a mode with negative spectral residue. This difference originates from the opposite sign of the Hartree feedback in the self-consistent mass renormalization, which enhances the effective ghost mass while reducing the effective mass of the healthy excitation.

\begin{figure}[h]
    \centering
    \includegraphics[width=1\linewidth]{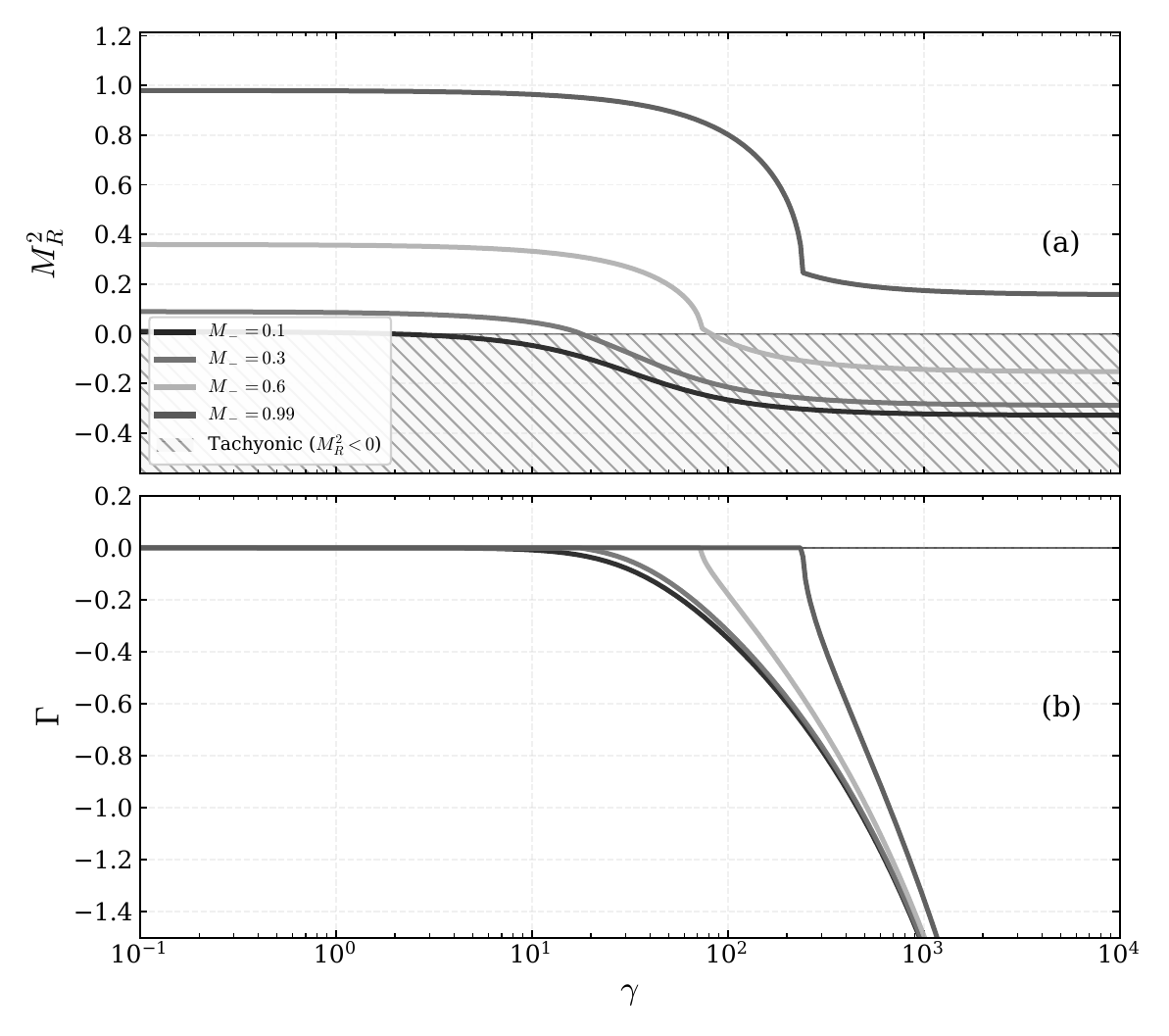}
    \caption{(a) Effective mass squared $M_R^2 $ of the healthy sector as a function of the dissipative coupling $\gamma $ (in units of the UV cutoff $\Lambda$), for $\Delta = 0.1$ and several values of the bare mass $M_- $. The mass decreases monotonically with increasing $\gamma$, and for a set of values of the bare mass ($M_-< \sqrt{2/3}\Lambda$) the effective mass vanishes at finite coupling, signaling a tachyonic instability. No bifurcation is observed, in contrast with the ghost sector. (b) Dissipative width $\Gamma $ of the healthy sector as a function of $\gamma $, for $\Delta = 0.1$ and different values of $M_- $. For most parameters, the width remains infinitesimally small, indicating the absence of a non-perturbative dissipative phase transition. Only when the effective mass approaches its minimum, a negative width emerges at sufficiently strong coupling. The absence of bifurcation and the sharp onset of $\Gamma \neq 0$ contrast with the ghost sector, where a clear bifurcation is present.}
    \label{fig3}
\end{figure}

\begin{figure}[h]
    \centering
    \includegraphics[width=1\linewidth]{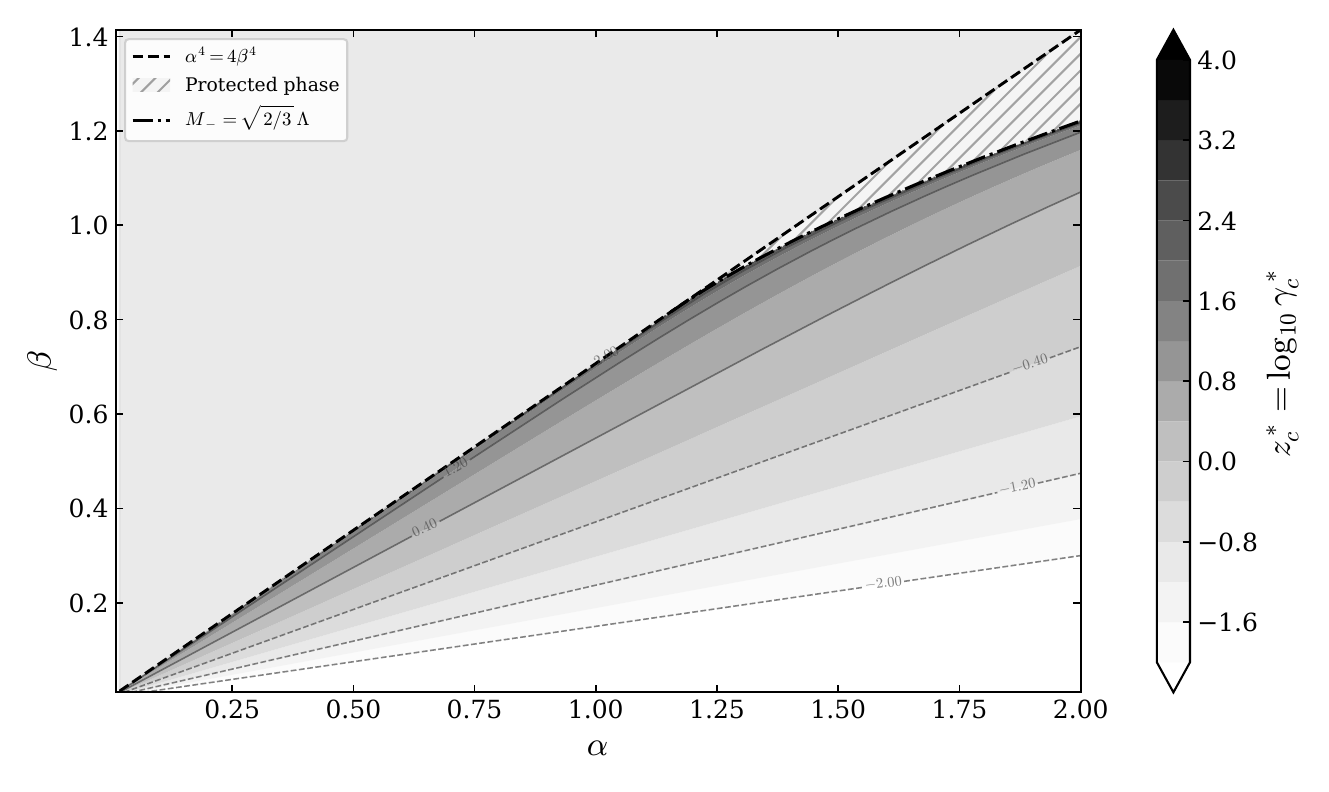}
    \caption{Phase diagram in the $(\alpha, \beta)$ plane showing the critical coupling $z_c^* = \log_{10} \gamma_c^*$ at which the effective mass of the healthy sector satisfies $M_R^2(\gamma_c) = 0$ (onset of the tachyonic regime). The dashed curve corresponds to the degeneracy boundary $\alpha^4 = 4\beta^4$ where $\Delta = 0$. The dash-dotted curve indicates the critical scale $M_- = \sqrt{2/3}\,\Lambda$, which marks the termination of the dissipative transition line. Below this curve, the system undergoes a tachyonic transition for a finite critical coupling $z_c^*$; above it, the transition ceases to exist within finite coupling. Contour lines indicate constant values of $z_c^*$. The shaded region corresponds to $\alpha^4 < 4\beta^4$, where the bare masses become complex.}
    \label{fig4}
\end{figure}

It is instructive to contrast the behavior of the ghost sector with that of a healthy sector coupled to an analogous dissipative environment via the jump operator $L \sim \Phi^2$. Solving the corresponding gap equations numerically (see Figs.\ref{fig2} and \ref{fig4}) reveals a markedly different picture: no bifurcation occurs for any masses $M_-$. 
In contrast with the ghost sector, the healthy branch does not develop multiple self-consistent dissipative solutions, indicating the absence of a non-equilibrium bifurcation structure.
We find that the effective mass $M_R$ decreases with the coupling constant $\gamma$, and for a critical value $M_R^2<0$, turning into a tachionic excitation. But, for massive enough bare mass $M_-$, the effective mass squared becomes positive for any finite coupling. In this regime, the bare mass scale dominates over the dissipative corrections, preventing the self-consistent flow toward the tachyonic sector. Looking to Fig \ref{fig3}, we see that the effective width $\Gamma$ is negative and becomes a non-perturbative effect only for a subtle value of the coupling constant $\gamma$. The small magnitude of the dissipative width indicates that the healthy excitation largely preserves its quasiparticle character throughout the evolution. Nonetheless, there is no bifurcation and the dissipative transition is smooth with respect to the coupling constant $\gamma$. The absence of branch splitting suggests that the healthy sector undergoes a continuous crossover toward the tachyonic regime rather than a genuine dissipative bifurcation. One observes in Fig. \ref{fig4} the emergence of a critical scale,
\begin{equation}
M_-^*=\sqrt{\frac{2}{3}}\,\Lambda,
\end{equation}
at which the critical dissipation diverges, $z_c\to\infty$, indicating that beyond this threshold the dissipative phase transition is no longer dynamically accessible within finite coupling. The appearance of the cutoff scale in the critical condition reflects the effective-field-theory character of the present dissipative description. The divergence of the critical coupling signals the termination of the dissipative transition line in parameter space. The region bounded by
\begin{equation}
\Lambda\left(
\frac{2}{3}\frac{\alpha^2}{\Lambda^2}-\frac{4}{9}\right)^{1/4}<\beta <
\frac{\alpha}{\sqrt2},
\end{equation}
corresponding to
\begin{equation}
M_->M_-^*~~,
\end{equation}
defines a protected phase in which the dissipative transition ceases to exist. Geometrically, this region corresponds to a disconnected sector of parameter space where the dissipative flow never reaches the critical surface associated with the tachyonic transition. In this regime, the effective mass remains positive, i.e., $M_R^2>0$ for any finite dissipation strength $\gamma$, indicating that the healthy-sector mass scale dominates over the dissipative corrections and prevents the emergence of the tachyonic phase.

Hence, the dynamical stabilization mechanism identified for the ghost is not a universal feature of quadratic Lindblad self-interactions, but rather a peculiar consequence of the negative spectral residue carried by the Ostrogradsky mode. The comparison between the two sectors, therefore, suggests that the dissipative stabilization mechanism is intrinsically tied to the ghost-like spectral structure of the higher-derivative theory.
%-----------------------------------------------------

%------------------------------------------------------------------
\section{Final Remarks}

In this work, we investigated how dissipative effects induced by the coupling to external baths modify the pole structure of ghost excitations in the Ostrogradsky model within a nonequilibrium Keldysh--Lindblad framework. We showed that, above a critical dissipative coupling $\gamma_c$, the system dynamically generates $\Gamma<0$, leading to a dissipative suppression of ghost modes in the out-of-equilibrium regime. The resulting stabilization is purely kinetic and spectral in character, rather than a restoration of microscopic unitarity or of the lower boundedness of the Hamiltonian. The self-consistent 2PI gap equations exhibit a nontrivial bifurcated structure beyond the Nambu--Jona-Lasinio--type branch, signaling the emergence of a genuine dissipative phase transition in the ghost sector.

The bifurcated regime admits two qualitatively distinct nonequilibrium phases. In the first one, the effective mass grows while the dissipative width remains comparatively small, producing a massive quasiparticle-like ghost mode with suppressed spectral propagation. In the second regime, the dissipative width dominates over the effective mass, strongly broadening the spectral function and destroying the quasiparticle character of the excitation through overdamped dynamics and rapid decoherence. From an entropic perspective, these two branches appear to correspond to different entropy-production channels, separating a coherent massive phase from a strongly dissipative overdamped phase.

Our analysis also revealed an important asymmetry between the ghost and healthy sectors. While dissipation dynamically stabilizes the ghost branch through mass generation and spectral broadening, the healthy sector tends instead toward a tachyonic regime as the dissipative coupling increases. The existence of a protected phase satisfying $M_- > \sqrt{2/3}\Lambda$ further indicates that the dissipative transition is not universally accessible in parameter space, but depends sensitively on the spectral structure of the theory.

Nonetheless, several important issues remain open. The validity of the Markov approximation, the dependence on the ultraviolet cutoff, the restriction to homogeneous configurations, and the dynamical stability of the bifurcated branches require further investigation. Extensions including interactions between the healthy and ghost sectors, non-Markovian effects, and generalizations to fields with spin are natural directions for future work. The present framework may also shed light on the role of ghost excitations in higher-derivative quantum gravity, nonequilibrium cosmology, and strongly interacting out-of-equilibrium quantum field theories.

An interesting possibility emerging from the present analysis concerns the formation of dissipatively induced bound states involving both the healthy and ghost sectors. While the present work considered a quadratic jump operator of the form $L\sim\tilde{\Phi}^2$, a bilinear Lindblad operator $L\sim\Phi\tilde{\Phi}$ would dynamically couple the two sectors already at the quadratic level through mixed Keldysh vertices. This opens the possibility of studying a Bethe--Salpeter equation for the composite operator $J=\Phi\tilde{\Phi}$, whose two-point function could develop a physical bound-state pole. Recent studies~\cite{Asorey26,Modesto16} suggest that ghost bound states may restore an effective notion of unitarity in higher-derivative theories. Whether a similar mechanism can emerge purely from dissipative dynamics remains an open and intriguing question that deserves further investigation.

%------------------------------------------------------------------

\section*{Acknowledgments}
 YMPG is supported by a postdoctoral
grant from Funda\c c\~ao Carlos Chagas Filho de Amparo \`a
Pesquisa do Estado do Rio de Janeiro (FAPERJ), Grant
No. E-26/200427/2025.
%%%%%%%%%%%%%----------------------------------------------------------------
\section*{Appendix I: Derivation of the Lindbladian Action from Microscopic Baths}\label{AppI}

In this appendix, we outline the derivation of the effective dissipative action employed throughout the main text. We consider the higher-derivative scalar field $\phi$ coupled to a set of independent environmental degrees of freedom described by $N$ massive scalar baths $B_i$, each one kept at equilibrium temperature $T_i$, with $i=1,\dots,N$. The system Lagrangian is
\begin{equation}
    \mathcal{L}_{\phi}
    =
    \phi\left(
    \square^2+\alpha^2\square+\beta^4
    \right)\phi .
\end{equation}

The baths are assumed to be weakly coupled, Gaussian, and sufficiently massive such that their intrinsic dynamics are much faster than the characteristic timescale of the subsystem. The Keldysh action for the bath fields is
\begin{equation}
    S_{B_i}
    =
    \int d^4x
    \left[
    \frac12(\partial_\mu B_{+,i})^2
    -
    \frac12 m_i^2 B_{+,i}^2
    -
    (B_+\rightarrow B_-)
    \right].
\end{equation}

The interaction between the subsystem and the baths is chosen quadratic in the ghost  field $\tilde{\Phi}$,
\begin{equation}
    S_{\mathrm{int}}
    =
    \sum_i g_i
    \int d^4x
    \left(
    \tilde{\Phi}_+^2 B_{+,i}
    -
    \tilde{\Phi}_-^2 B_{-,i}
    \right),
\end{equation}
where $g_i$ are small coupling constants. We assume the weak-coupling regime such that the subsystem does not thermalize with the environment, allowing the bath dynamics to be integrated out perturbatively.

Introducing the Keldysh rotation,
\begin{equation}
    \tilde{\Phi}_c
    =
    \frac{1}{\sqrt2}(\tilde{\Phi}_+ + \tilde{\Phi}_-),
    \qquad
    \tilde{\Phi}_q
    =
    \frac{1}{\sqrt2}( \tilde{\Phi}_+ - \tilde{\Phi}_-),
\end{equation}
and similarly for the bath fields, the Gaussian path integral over $B_\pm$ can be performed exactly. The resulting functional integration generates nonlocal quartic interactions for the subsystem field:
\begin{eqnarray}\nonumber
&&S_{\mathrm{eff}}[\tilde{\Phi}_c,\tilde{\Phi}_q]
=
S_{\tilde{\Phi}}[\tilde{\Phi}_c,\tilde{\Phi}_q]
-
\frac12
\sum_i g_i^2
\int d^4x\,d^4y
\Big[
\\ \nonumber
&&\hspace{1.2cm}
4\,
\tilde{\Phi}_c(x)\tilde{\Phi}_q(x)
D_{K,i}(x-y)
\tilde{\Phi}_c(y)\tilde{\Phi}_q(y)
\\ \nonumber
&&\hspace{1.2cm}
+
2\,
\big(
\tilde{\Phi}_c^2(x)+\tilde{\Phi}_q^2(x)
\big)
\tilde{\Phi}_c(y)\tilde{\Phi}_q(y)
\\
&&\hspace{1.2cm}
\times
\big(
D_{R,i}(x-y)
+
D_{A,i}(y-x)
\big)
\Big].
\end{eqnarray}

The first contribution is governed by the Keldysh propagator of the baths and encodes fluctuations and dissipative noise, while the second contribution arises from the retarded and advanced propagators and corresponds to Hermitian corrections to the effective dynamics.

The bath correlators can be written in terms of their spectral densities,
\begin{equation}
    D_{K,i}(t-t')
    \sim
    \int d\omega\,
    J_i(\omega)
    \coth\left(\frac{\omega}{2T_i}\right)
    e^{-i\omega(t-t')},
\end{equation}
where $J_i(\omega)$ denotes the spectral function of the $i$-th bath. In the heavy-bath and high-temperature regime, the bath correlation time becomes much shorter than the characteristic timescale of the subsystem. Under the Markov approximation, the correlators become effectively local,
\begin{equation}
    D_{K,i}(x-y)
    \approx
    \widetilde D_{K,i}(0)\,
    \delta^{(4)}(x-y)
    \equiv
    2i\gamma_i
    \delta^{(4)}(x-y),
\end{equation}
with
\begin{equation}
    \gamma_i
    \propto
    \frac{T_i}{m_i^2}.
\end{equation}

Summing over all baths, the dissipative coefficient becomes
\begin{equation}
    \lambda_1
    =
    \sum_i
    \frac{4g_i^2T_i}{m_i^2}.
\end{equation}

On the other hand, the Hermitian corrections generated by the retarded propagators scale as
\begin{equation}
    \mathrm{Re}\,D_{R,i}
    \propto
    \frac{1}{m_i^2},
\end{equation}
such that their contribution,
\begin{equation}
    \widetilde\lambda_1
    \propto
    \sum_i
    \frac{g_i^2}{m_i^2},
\end{equation}
is parametrically suppressed in the high-temperature regime, $\tilde\lambda_1 \ll\lambda_1$. Therefore, the dominant contribution to the effective action is the local dissipative quartic vertex
\begin{equation}
    S_{\mathrm{diss}}
    \sim
    i\lambda_1
    \int d^4x\,
    \tilde{\Phi}_c^2\tilde{\Phi}_q^2~~,
\end{equation}
which possesses precisely the Lindblad structure employed in Eq.~\eqref{ghostself}. The corresponding imaginary contribution to the effective action generates the dissipative self-energy and the effective spectral broadening analyzed throughout the present work.

%========================================================
\section*{Appendix II: Exact Evaluation of the Gap Integral}\label{AppII}
The integral to be evaluated is:
\begin{equation}
    I(\Delta, \rho) = \int \frac{d^4p}{(2\pi)^4} \, \frac{\Delta}{p^2 - M_+^2 - 4\gamma\rho + 8i\gamma\tau}.
\end{equation}
After Wick rotation ($p_0 \to i p_{E0}$, $p^2 \to -p_E^2$), it becomes a Euclidean integral:
\begin{equation}
    I(\Delta, \rho) = -i\Delta \int \frac{d^4p_E}{(2\pi)^4} \, \frac{1}{p_E^2 + M^2},
\end{equation}
with the complex mass parameter:
\begin{equation}
    M^2 = M_R^2 + i M_I^2, \quad M_R^2 = M_+^2 + 4\gamma\rho, \quad M_I^2 = -8\gamma\tau.
\end{equation}

Introducing a sharp ultraviolet cutoff $\Lambda$ on the Euclidean momentum, the integral evaluates to:
\begin{equation}
    \int_0^\Lambda \frac{d^4p_E}{(2\pi)^4} \, \frac{1}{p_E^2 + M^2} 
    = \frac{1}{16\pi^2} \left[ \Lambda^2 - M^2 \ln\left( \frac{\Lambda^2 + M^2}{M^2} \right) \right].
\end{equation}
Therefore, the exact result for $I$ is:
\begin{equation}
    I(\Delta, \rho) = - \frac{i\Delta}{16\pi^2} \left[ \Lambda^2 - M^2 \ln\left( \frac{\Lambda^2 + M^2}{M^2} \right) \right].
\end{equation}

Writing $M^2 = M_R^2 + i M_I^2$, the logarithm expands as:
\begin{equation}
    \ln\left( \frac{\Lambda^2 + M^2}{M^2} \right) = \ln L + i \theta,
\end{equation}
with:
\begin{align}
    L &= \sqrt{ \frac{ (\Lambda^2 + M_R^2)^2 + (M_I^2)^2 }{ (M_R^2)^2 + (M_I^2)^2 } }, \\
    \theta &= \arctan\left( \frac{M_I^2}{\Lambda^2 + M_R^2} \right) - \arctan\left( \frac{M_I^2}{M_R^2} \right).
\end{align}
The real and imaginary parts of $I$ are then:
\begin{align}
    \operatorname{Re} I &= \frac{\Delta}{16\pi^2} \left[ M_I^2 \ln L + M_R^2 \, \theta \right], \\
    \operatorname{Im} I &= - \frac{\Delta}{16\pi^2} \left[ \Lambda^2 - M_R^2 \ln L + M_I^2 \, \theta \right].
\end{align}
%------------------------------------------------

%%%%%%%%%%%%%%%%%%%%%%%%%%%%%%%%%%%%%%%%%%%%%%%%%%%%%%%%%%%%%%%%%%%%%%%%%%%%%%%%%%%%%%%%%%%%%%%%%%%%%%%5
%%%%%%%%%%%%%%%%%%%%%%%%%%%%%%%%%%%%%%%%%%%%%%%%%%%%%%%%%%%%%%%%%%%%%%%%%%%%%%%%%%%%

\end{document}